\documentclass[final,5p,times,twocolumn,authoryear]{elsarticle}

\usepackage{amssymb}
\usepackage{lipsum}
\usepackage{hyperref}
\usepackage{dirtree}
\usepackage{longtable}
\usepackage{geometry}
\usepackage{pdflscape}
\usepackage{xcolor}
\usepackage{listings}

\definecolor{codegreen}{rgb}{0,0.6,0}
\definecolor{codegray}{rgb}{0.5,0.5,0.5}
\definecolor{codepurple}{rgb}{0.58,0,0.82}
\definecolor{backcolour}{rgb}{0.95,0.95,0.92}

\lstdefinestyle{mystyle}{
    backgroundcolor=\color{backcolour},   
    commentstyle=\color{codegreen},
    keywordstyle=\color{magenta},
    numberstyle=\tiny\color{codegray},
    stringstyle=\color{codepurple},
    basicstyle=\ttfamily\small,
    breakatwhitespace=false,         
    breaklines=true,                 
    captionpos=b,                    
    keepspaces=true,                              
    numbersep=5pt,                  
    showspaces=false,                
    showstringspaces=false,
    showtabs=false,                  
    tabsize=2
}
\newcommand{\inline}[1]{\fboxsep2pt
\colorbox{backcolour}{\lstinline|#1|}}

\journal{Astronomy $\&$ Computing.}
\myfooter[L]{\scriptsize
  \begin{tabular}[b]{l}
  \small Preprint submitted to Astronomy $\&$ Computing. \\
 \scriptsize \textcopyright~2025. This manuscript version is made available under the \href{https://creativecommons.org/licenses/by-nc-nd/4.0/}{CC-BY-NC-ND 4.0 license}.\\ 
  \end{tabular}}

\newcommand{\EMC}{\texttt{Exo-MerCat}}
\newcommand{\EMCv}[1]{\texttt{Exo-MerCat} v{#1}}

\begin{document}

\begin{frontmatter}



\title{\EMCv{2.0.0}: updates and open-source release of the Exoplanet Merged Catalog {software}}

\author[nasa,eth]{Eleonora Alei\fnref{label1}}
\ead{eleonora.alei@nasa.gov}
\fntext[label1]{NPP Fellow}

\author[oar,asissdc]{Silvia Marinoni}
\author[oats]{Andrea Bignamini}
\author[oapd,roma3]{Riccardo Claudi}

\author[oats]{Marco Molinaro}
\author[oar]{Martina Vicinanza}
\author[oapa]{Serena Benatti}
\author[oato]{Ilaria Carleo}
\author[nasa]{Avi Mandell}
\author[eth]{Franziska Menti}

\author[asi,asissdc]{Angelo Zinzi}

\affiliation[nasa]{organization={NASA Goddard Space Flight Center},
           addressline={8800 Goddard Rd}, 
           city={Greenbelt},
           postcode={20771}, 
           state={MD},
           country={USA}}
                    
\affiliation[eth]{organization={ETH Zurich, Inst. f. Teilchen- und Astrophysik},
           addressline={Wolfgang-Pauli-Str. 27}, 
           city={Zurich},
           postcode={8093},
           country={Switzerland}} 
\affiliation[oar]{organization={INAF Osservatorio Astronomico di Roma},
           addressline={via Frascati, 33}, 
           city={Monte Porzio Catone (Roma)},
           postcode={00078}, 
           country={Italy}}

\affiliation[asissdc]{organization={Space Science Data Center-ASI},
           addressline={via del Politecnico SNC}, 
           city={Roma},
           postcode={00133}, 
           country={Italy}}

\affiliation[oats]{organization={INAF Osservatorio Astronomico di Trieste},
           addressline={via G.B. Tiepolo, 11}, 
           city={Trieste},
           postcode={34143}, 
           country={Italy}}
\affiliation[oapd]{organization={Osservatorio Astronomico di Padova},
           addressline={vicolo Osservatorio, 5}, 
           city={Padova},
           postcode={35122}, 
           country={Italy}}        
  \affiliation[roma3]{organization={Mathematic \& Physics Dep. University of Roma Tre},
           addressline={via della vasca navale, 84}, 
           city={Roma},
           postcode={00146}, 
           country={Italy}}        

\affiliation[oapa]{organization={INAF Osservatorio Astronomico di Palermo},
           addressline={Piazza del Parlamento 1}, 
           city={Palermo},
           postcode={90134},
           country={Italy}} 
 \affiliation[oato]{organization={INAF Osservatorio Astronomico di Torino},
           addressline={Via Osservatorio 20}, 
           city={Pino Torinese},
           postcode={10025},
           country={Italy}} 
 \affiliation[asi]{organization={Agenzia Spaziale Italiana},
           addressline={via del Politecnico SNC}, 
           city={Roma},
           postcode={00133}, 
           country={Italy}}
               
% \author[1]{Alei, Eleonora},\author[2]{Marinoni, Silvia}
% \affiliation[1]{organization={NASA Goddard Space Flight Center},%Department and Organization
%             city={Earth},
% \affiliation[2]{organization={ASI},%Department and Organization
%             city={Earth},

\begin{abstract}
Exoplanet research is at the forefront of contemporary astronomy recommendations. As more and more exoplanets are discovered and vetted, databases and catalogs are built to collect information. Various resources are available to scientists for this purpose, though every one of them has different scopes and notations. In \citet{Alei2020} we described \EMC, a {script} that collects information from multiple sources and creates a {homogenized} table. In this manuscript, we announce the release of {the} \EMCv{2.0.0} {script} as an upgraded, tested, documented and open-source {software to produce catalogs}. The main upgrades on the script concern: 1) the addition of the TESS Input Catalog and the K2 Input Catalog as input sources; 2) the optimization of the main identifier queries; 3) a more complex merging of the entries from the input sources into the final catalog; 4) some quality-of-life improvements such as informative flags, more user-friendly column headers, and log files; 5) the refactoring of the code in modules. We compare the performance of \EMCv{2.0.0} with the previous version and notice a substantial improvement in the completeness of the sample, thanks to the addition of new input sources, and its accuracy, because of the optimization of the script.
\end{abstract}




\end{frontmatter}


\section{Introduction}
\label{introduction}
The last few years have seen a huge advancement in the detection and characterization of extrasolar planets. From the discovery of the first exoplanet in 1995 \citep{mayorqueloz1995}, we discovered more than 7000 candidates and 6000 confirmed exoplanets in our galactic neighborhood -- and these numbers increase daily. A variety of space- and ground-based telescopes have discovered a large number of planetary candidates, and more facilities will be doing so in the near- and long-term future. As ancillary resources for any researcher and to store data in a {homogenized} way, catalogs are developed and maintained. Such resources serve both as input target lists for mission scheduling, as well as output storage from past observations. 

In 2020, we released the first version of \EMC\ \citep{Alei2020}, {a Python script that produces a catalog of known and candidate exoplanets by merging} data collected from multiple input sources. As described in \cite{Alei2020}, merging entries belonging to different source catalogs proved to be a challenging problem because of the presence of aliases (i.e., the same exoplanet listed in one or more catalogs using different nomenclature). For this reason, \EMC\ retrieves a common name for the planet target, linking its host star name with the preferred identifier in the most well-known stellar databases. This makes the merging of different input sources possible and it has allowed the {creation} of a {homogenized} catalog containing data from the four major input catalogs used in the community: the {\it Exoplanets Encyclopaedia}\footnote{\url{http://exoplanet.eu/}} \citep{Schneider2011},  the {\it NASA Exoplanet Archive}\footnote{ \url{https://exoplanetarchive.ipac.caltech.edu/}} \citep{Akeson2013}, the {\it Open Exoplanet Catalogue}\footnote{\url{https://openexoplanetcatalogue.com}}  \citep{Open} and the {\it Exoplanet Orbit Database} catalog\footnote{\url{http://exoplanets.org/}} \citep{Wright2011}.
Because of the nature of the merge, this catalog is not self-consistent (i.e., all measurements for the same target do not belong to the same reference paper). Rather, it is a collection of the most precise estimates for each planetary parameter (mass, minimum mass, radius, period, semi-major axis, eccentricity, inclination), based on the lowest relative error, and the corresponding reference for each preferred estimate. We refer to \cite{Alei2020} for further details on the implementation of \EMCv{1.0.0}.

{The catalog produced by \EMC} has been used as a source for input target lists for space-based missions, as well as an additional tool for various publications (see Sect. \ref{sec:value}). In time, we received specific requests to update the code to accommodate the needs of the current users, which we deemed useful and worthy of a software update.
Furthermore, the input sources have undergone significant changes (e.g., the migration of the NASA Exoplanet Archive to a new table and notation), which made the ingestion into \EMC~fail at runtime.  Finally, to streamline software updates and to allow continuous maintenance of the code, we decided to release the updated version of \EMC\ as open-source code on GitHub. 

In this paper, we present \EMCv{2.0.0} as a more optimized, robust, and user-friendly version of the Exoplanet Merged Catalog {software}. 
In Sect. \ref{sec:updates} we describe the updates that were performed on \EMC. In Sect. \ref{sec:discussion} we compare the performance of v2.0.0 compared to the previous version, we discuss the value of such a resource for the community, and we describe the open-source philosophy and maintenance of the {script} going forward. We provide a summary of the work in Section \ref{sec:conclusion}.

\section{\EMCv{2.0.0}: main features}\label{sec:updates}

 Every version of \EMC~fulfills the following tasks:
\begin{enumerate}
    \item Ingesting the input sources as tables. Since different input sources come in different formats, the script can flexibly download the tables and convert them in the same \texttt{.csv} format.
    \item {Homogenizing} the input sources to handle the same type of data, in terms of table columns and units. Different input sources have different data available in various formats and are made as similar as possible to ease the future tasks.
    \item Selecting a host star identifier for each target in the tables that is the same across the input sources. This is done by querying the host star with stellar catalogs.% such as SIMBAD and the TESS Input Catalog.
    \item Merging the entries from the various input sources into a single entry. When multiple estimates for the same planetary parameter are available, \EMC~prefers the most precise measurement. 
\end{enumerate}

While this has not changed between the various versions, in \EMCv{2.0.0} these tasks are executed in a more {streamlined} and optimized way. We summarize the main differences between \EMCv{2.0.0} and its previous version in Table \ref{tab:updates}. Each process in the itemized list and in the table is described in more detail in Sections \ref{sec:ingestion} to \ref{sec:outputs}.

{The \EMC~script is written in Python and it has been tested for Python v3.8 and above.}

% If downloaded and installed locally, the script can be launched with the following command:

% \begin{lstlisting}[language=Python]
% exomercat [-h] [-v] [-w] [-l] [-d DATE] function
% \end{lstlisting}

% The user can select optional arguments: \inline{-h} (or \inline{--help}) to print a help message; \inline{-v} (or \inline{--verbose}) to increase the verbosity of the output (see Sect. \ref{sec:outputs}), \inline{-w} (or \inline{--warnings}) to show warnings; \inline{-l} (or \inline{--local}) to load the most recent {homogenized} catalogs (see Sect. \ref{sec:homogenization}); \inline{-d YYYY-MM-DD} (or \inline{--date YYYY-MM-DD}) to load the input sources at a specific date in YYYY-MM-DD format (see Sect. \ref{sec:ingestion}). Possible functions to be run are: \inline{input}, which executes the download of the input sources and their homogenization (see Sect. \ref{sec:ingestion} and \ref{sec:homogenization}); \inline{run}, which joins the input sources to generate the \EMC~catalog (see Sect. \ref{sec:mainid} and \ref{sec:merging}); \inline{check}, which performs sanity checks on the output; \inline{tests} which performs unit tests (see Sect. \ref{sec:opensource}).

\renewcommand*{\arraystretch}{1.4}

\begin{table*}[htpb]
    \centering
    \begin{tabular}{p{.45\linewidth}|p{.45\linewidth}}
    \hline\hline

        \EMCv{1.0.0} & \EMCv{2.0.0} \\\hline
                     \multicolumn{2}{c}{Input sources ingestion (Sect. \ref{sec:ingestion})}\\
         \hline
         \hangindent=1em Hard-coded links to sources &  \hangindent=1em User-specified links to sources (to be configured in \inline{input\_sources.ini}) 
\\
         \hangindent=1em Deprecated links to input sources&  \hangindent=1em Updated input sources download
\\
         \hangindent=1em Exoplanet Orbit Database as source&  \hangindent=1em Removed Exoplanet Orbit Database from input sources
\\
         -- & \hangindent=1em  Option to load specific versions of the catalogs when available in the \inline{InputSources/} folder
\\\hline
\multicolumn{2}{c}{Input sources {homogenization} (Sect. \ref{sec:homogenization})}\\
         \hline
        
\hangindent=1em Default .0x to letter replacements (.01=b,.02=c…)& \hangindent=1em{\textit{$^\star$performed later}}
\\
\hangindent=1em  Potential Brown Dwarf: set letter value as ``b''& \hangindent=1em Potential Brown Dwarf: set letter value as ``BD''
\\
       \hangindent=1em  Hard-coded replacements& \hangindent=1em User-specified replacements
\\

\hangindent=1em KOI look-up table: replace updated status &\hangindent=1em KOI look-up table: save both old and updated status \\

 % - & \hangindent=1em Meaningful logging: writing non-used replacements for optimization
\hline

                     \multicolumn{2}{c}{Host star identifier selection (Sect. \ref{sec:mainid})}\\
         \hline
% - &\hangindent=1em  Comparison on binary systems to infer information on \inline{binary} column\\
 \hangindent=1em Check that coordinates match within 36 arcsec; replace mismatches with mode or mean &\hangindent=1em  Check that coordinates match within 1 arcsec; log disagreements, but no replacement applied
\\

 \hangindent=1em Main Identifier: Query by host in SIMBAD. If not found,  query by coordinates at increasing radii (up to $\approx$10 arcsec)  in SIMBAD, KIC, K2, Gaia until all are found & \hangindent=1em Main Identifier: Query by host in SIMBAD and TIC. If not found, query by coordinates at increasing radii (up to 1 arcsec) on SIMBAD and TIC. If not found, use catalog identifier
\\
 Check duplicated entries & {\textit{$^\bullet$performed later}}
\\\hline
 \multicolumn{2}{c}{Merging the entries (Sect. \ref{sec:merging})}\\\hline

{\textit{$^\star$performed earlier}}& \hangindent=1em Check for inconsistent \inline{letter} values by grouping by (\inline{main\_id}, \inline{binary} and \inline{p} or \inline{a}); replace ``.0d'' into letter when possible  and check inconsistent letters
\\
 \hangindent=1em Group entries on (\inline{main\_id}, \inline{binary} and \inline{letter}) for merging & \hangindent=1em  Group entries on (\inline{main\_id}, \inline{binary},\inline{letter}) for merging, then divide in subgroups by \inline{p} or \inline{a}  %Cases: 1. if period is the same, merge into single entry; 2. if contrasting periods, merge separately by period value; 3. if no period, check semi-major axis: 3a. If same semi-major axis, merge into single entry; 3b. if different semi-major axes, merge by semi-major axis value. 4. no period and no semi-major axis available, fallback and merge all together and write in log file
\\
{\textit{$^\bullet$performed earlier}}& \hangindent=1em Check for duplicated values in merging entries
\\
\hangindent=1em  If status is not consistent, prefer FALSE POSITIVE or CANDIDATE if at least one is present&\hangindent=1em  If status is not consistent, write CONTROVERSIAL.
\\
Alphanumeric string to represent status of merged entries & User-friendly status string
\\
 Catalog name selected from available aliases& {Common naming convention}: \inline{main\_id} + \inline{binary} + \inline{letter}
\\
-- & \hangindent=1em Use of flags to highlight errors in the merging process for easy filtering \\\hline
  \multicolumn{2}{c}{Outputs (Sect. \ref{sec:outputs})}\\\hline

No brown dwarf filtering& \hangindent=1em Possible filtering of brown dwarfs based on mass criterion ($m$ or $m\sin{i}<20 M_{Jup}$) 
\\
 \hangindent=1em Logging provided only at run-time & \hangindent=1em  Logging provided at run-time (if desired), plus saved in ancillary files for asynchronous reference
\\\hline

    \end{tabular}
    \caption{Summary of the main differences between \EMCv{1.0.0} and \EMCv{2.0.0}. {The star and bullet symbols link the tasks that are performed at different points in the script in the two versions.}}
    \label{tab:updates}
\end{table*}

\subsection{Input sources ingestion}\label{sec:ingestion}

\EMCv{2.0.0} takes into account the following input sources: the NASA Exoplanet Archive (hereafter NASA); the Exoplanet Encyclopaedia (hereafter EU); the Open Exoplanet Catalogue (hereafter OEC); the EPIC/K2 Planets and Candidates Table\footnote{\url{https://exoplanetarchive.ipac.caltech.edu/cgi-bin/TblView/nph-tblView?app=ExoTbls&config=k2pandc}} (hereafter EPIC); and the TESS Project Candidates Table\footnote{\url{https://exoplanetarchive.ipac.caltech.edu/cgi-bin/TblView/nph-tblView?app=ExoTbls&config=TOI}} (hereafter TOI). Like in \EMCv{1.0.0}, the Kepler Objects of Interest Table\footnote{\url{https://exoplanetarchive.ipac.caltech.edu/cgi-bin/TblView/nph-tblView?app=ExoTbls&config=cumulative}} (hereafter KOI) is used as a look-up table to check and {homogenize} the status of the Kepler targets. 

In this new version of the software, we kept three core catalogs from v1.0.0 (NASA, EU, OEC).  We removed the Exoplanet Orbit Database from the input sources, since the table has been retired, and additional tables were either added from scratch (in the case of the TOI table) or promoted from look-up table to input sources (in the case of EPIC). We chose to add or promote these tables since they are frequently updated and contain the latest updates on the TESS mission results and on further analyses on K2 data. The Kepler Objects of Interest table has not been updated since 2018, as all vetting has been completed. However, the KOI table is still useful as a look-up table since it reports all the updated statuses of each target in the Kepler sample and it is able to fix discrepancies in the other catalogs. 

Some of the input tables changed between \EMCv{1.0.0} and the current version. The NASA Exoplanet Archive retired the tables that were used by \EMCv{1.0.0} for the NASA, KOI, and EPIC; new tables are hosted now on different URLs. Some column names were replaced and other columns were removed/joined together. Furthermore, the NASA catalog (the \texttt{pscomppars} table) now contains heterogeneous data from different sources. For more details, we refer the reader to the NASA Exoplanet Archive website\footnote{\url{https://exoplanetarchive.ipac.caltech.edu/docs/transition.html}}. Similarly, the EU catalog has undergone major updates in their database structure and layout; some column names were changed and the relative URL was modified.
These updates required an adjustment of the input source ingestion to ensure that the \EMC\ script would not fail at runtime. The changes consisted in updating the URLs and the naming conventions according to the documentation provided by the NASA Exoplanet Archive and the Exoplanet Encyclopaedia.

Finally, the Open Exoplanet Catalogue now provides a cumulative \texttt{.xml} file in addition to the separate \texttt{.xml} files for each system. \EMC~has been updated to download the cumulative file instead.

In \EMCv{1.0.0}, the URLs that allowed the download of the input source tables were hard-coded. In v2.0.0, \EMC~reads the \inline{input\_sources.ini} file to get this information, as well as the destination and name of the downloaded file. The user can update the links to the source files by editing the \inline{input\_sources.ini} file. In this way, it will be possible to quickly adapt to changes in URLs and addition/removal of specific columns, or input catalogs directly.

By default, the catalogs are downloaded directly into the \inline{InputSources} folder. The file name includes the download date in \inline{YYYY-MM-DD} (month, day, year) format.{ When running the script, the user can specify the flag \inline{-date YYYY-MM-DD} (or \inline{--d YYYY-MM-DD}) to load the version of the input catalogs downloaded at a specific date, provided they are present in the \inline{InputSources} folder. If the download of a current version of any of the six tables fails at runtime, the run will continue provided that previous tables are present in the folder. In this case, the script automatically prints a warning for the user and loads the most recent table available. If no earlier tables are available, the script fails at runtime. Issues on the download of the input catalogs are often temporary and will likely be fixed at the source in a short amount of time.
}
\subsection{Input sources {homogenization}}\label{sec:homogenization}

Each input source collects specific datasets for a selected amount of targets. The choice of including or excluding a specific target depends on the philosophy of each catalog: for example, the NASA catalog only displays confirmed planets; the EU catalog collects also candidate and free-floating planets, as well as Solar System objects and trojans. Various catalogs could also adopt different mass cutoffs when it comes to the distinction between exoplanets and brown dwarfs \citep[see][for details]{Alei2020}. 
We cannot expect to have the same information among different catalogs. Still, the format of each table can be {homogenized} to allow a fairer comparison and merging of the entries that are included in two or more catalogs.

The \EMC~script performs specific operations for each input catalog to ensure that the {homogenized} output can easily allow for a comparison among the various tables. 
First, it selects and renames all the columns in the original files that will be used for the merging. These include: the planet name and its host star; the aliases by which the planet and/or the star are known; the planet letter and the binary string associated to the host star; the discovery method and year; and the estimates (including uncertainties and reference) for planetary mass, minimum mass, planetary radius, semi-major axis, eccentricity, period, and inclination. 

{We do not include stellar parameters or radial velocity and transit observables since those are not uniformly stored in all input sources, or rely on self-consistent measurements that cannot be provided in this joint catalog per definition.  For these parameters, we recommend querying specific stellar or planetary catalogs.
}

After ingestion, the planet and star names are {homogenized} to have a similar notation in the various catalogs. Normally, a confirmed exoplanet is known by the name of its host star followed by a lowercase letter (most commonly, ``b'' for single-planet systems). When there are multiple planets orbiting the same host star, subsequent letters from ``b'' to ``j''\footnote{As of October 25, 2024, there is only one occurrence of a letter  ``j'', for the candidate planet HD 10180 j.} are used to label all the planets in the system, by year of discovery. Candidate exoplanets, generally revealed by transit detection space missions,  are instead labeled as the name of their host star in the input catalog of the mission followed by the string ``.0d'' where ``d'' represents an integer that starts from 1 and increases as more candidates are found in the same planetary system. These candidates, once vetted by complementary methods, are then confirmed or rejected. In case of confirmation, they then become known with their planetary letter. 

In \EMCv{1.0.0} the candidate ``.0d'' indexing was replaced by the corresponding letter (assuming that ``.01'' becomes ``b'', ``.02'' becomes ``c''...). However, in reality this relationship is not so obvious, as the vetting process might take more or less time depending on the candidate planet, and automatically transforming the candidate letter into the confirmed letter might have lead to mismatches in the merges of a few candidates. It might have also been a source of confusion for the user, since candidate planets are not normally found labeled with a letter, but only with ``.0d'' strings. In the new version of the code, we removed this automatic replacement between numeric string and letter while {homogenizing} the catalogs to avoid unintentional mismatches. This choice leads to a subsequent issue: the same confirmed planet might have been present in different catalogs with its candidate name (ending in ``.0d'') as well as its official name (with the corresponding letter); we resolve these differences when collapsing the duplicated catalog entries into the merged \EMC~entry (see Sect. \ref{sec:merging}). 

In \EMCv{1.0.0}, we also automatically added the letter ``b'' to all entries that did not show a lowercase letter nor a ``.0d'' string in the name. This is often the case for substellar objects. In \EMCv{2.0.0}, we reserve the specific value ``BD'' of the planet letter column (called \inline{letter}) for all entries that do not show a letter nor a ``.0d'' string in their name. It is thus possible to filter these candidates out easily from the final catalog.

In the case of binaries, the information content in the input sources is not uniform. Specifically, OEC provides a flag that establishes if the planet orbits in a P-type (circumbinary) or S-type orbit (around one of the binary companions), but it is not often clear about which companion hosts the planet. In \EMCv{1.0.0}, the \inline{binary} value that was only identified as ``S-type'' was discarded from the {homogenized} output. In \EMCv{2.0.0}, this partial information is still retained and, when possible, corrected at a later stage (see Sect. \ref{sec:mainid}).

As was done in v1.0.0, the \EMC~script checks that the status associated with the entries in the input catalogs is up-to-date. This can easily be done for the Kepler targets (the majority of the confirmed planets known so far) by comparing the value of the status of each Kepler target included in the input sources with the status in the KOI table. This is relevant especially in the case of candidates that were then revealed as false positives. While the status might not have been updated in some input sources we consider in \EMC, it has been updated in the KOI table, so we refer to that as the most accurate status. As of v2.0.0, both the original status and the updated status after the check on the KOI table is stored in \EMC~in the \inline{original\_status\_string} and \inline{checked\_status\_string} columns.

In \EMCv{1.0.0} we implemented hard-coded modifications to some specific targets in the sample that appeared to possess errors in their input catalogs, or to manually {homogenize} names of special targets (e.g., the planets in the PSR 1257+12 system, known with the suffixes ``A'', ``B'', and ``C'', normally used for stellar binary companions). These corrections were performed at various points within the script. This put an additional layer of complexity for any user wanting to customize such replacements. As of v2.0.0, any user  can force the replacement of a specific cell value (name, host star, binary letter, or coordinates) by providing the \inline{replacements.ini} file. This action might be necessary to: 1) replace known mistakes on names of planets or host stars that have been known to occur in the input catalogs; 2) drop selected elements from the list (e.g., Solar System items, which are also included in the EU catalog); 3) force a specific value of the binary suffix (see Sect. \ref{sec:homogenization}); 4) manually set specific right ascension and declination values. The \inline{replacements.ini} file will surely change with version updates, as the input sources evolve. Users are invited to modify and submit any additional replacement to be performed via push request (see Sect. \ref{sec:opensource}). 

The log files that are produced at run time provide information on any additional replacement to be performed or if any replacement is no longer necessary. In the latter case, the \inline{replace\_known\_mistakes.txt} log file lists all the replacements that were not used when {homogenizing} each source. We expect to always have some entries in this file, since some replacements are specifically tailored for one source catalog and might not be necessary for all the others. However, the user could optimize the contents of \inline{replacements.ini} if no replacement is performed in any source. Some replacements could also no longer be necessary if the issue was fixed by the source catalog at some point in time.

{In all versions of \EMC, the coordinates are converted to decimal degrees for KOI and OEC. Furthermore, theoretical estimates of masses and radii are removed. In \EMCv{1.0.0} all parameter estimates that had infinite or null errors were also removed. This is no longer the case for \EMCv{2.0.0}, which retains partial information when available.}

Separate files for the {homogenized} entries for each input catalog are then collected in the  \inline{StandardizedSources} folder as \texttt{.csv} files.

\subsection{Host star identifier selection}\label{sec:mainid}

In every version of \EMC, once the input catalogs have been {homogenized}, they are concatenated in a joint \texttt{pandas} DataFrame.
In principle, the same target can appear in the concatenated DataFrame up to five times, depending on how many input sources contained it. Since the same planet can be labeled differently depending on naming conventions and host star aliases, \EMC~performs various operations to assign a single identifier to each host star, to ensure a successful merging of each entry (see Sect. \ref{sec:merging}). 

All versions of \EMC~perform a check to identify matches between the \inline{host} column and the known aliases provided by each input catalog. In \EMCv{2.0.0}, the \inline{alias\_as\_host.txt} log file lists all of the host stars found by another name within the various input catalogs. 

In \EMCv{1.0.0}, all entries that shared the same host star were checked for coordinate mismatches. The DataFrame was grouped by the host star name; any coordinate that differed more than 0.01 degrees for the same host star was flagged and the coordinates of all entries of the group were changed to the mode value of the disagreeing coordinate. In \EMCv{2.0.0} we perform this test only on smaller fractions of the sample, such as planets orbiting binaries or planets whose main identifier was not found through name search (see below). In every coordinate check, the tolerance is by default smaller (1 arcsec) and the script does not perform any change but simply prints the name of the system in a log file for user vetting (see Sect. \ref{sec:outputs}). 

Binaries are handled in a slightly different way in \EMCv{2.0.0}. Previously, duplicated entries were grouped by host star name and more specific values of  \inline{binary} were replaced when more information was available (though excluding ``S-type'' binaries, see Sect. \ref{sec:homogenization}). \EMCv{2.0.0} groups by host star name and letter and allows the treatment of ``S-type'' binaries: ``S-type'' and null \inline{binary} values are replaced when additional information is present; if only null or ``S-type'' strings are present, the ``S-type'' value is preferred since it provides at least partial information. 

{Two flags are now introduced to highlight the entries that failed binary-related tests. A first coordinate check is performed for all binary systems (see above). Entries that fail the coordinate check will have the value of the \inline{binary\_coordinate\_mismatch\_flag} column set to \texttt{1}. If there are complex binary systems (e.g., coexistence of ``A'', ``B'', and/or ``AB'' binary values for the same \inline{host+letter} group), the value of the \inline{binary\_complex\_system\_flag} column is assigned to be equal to  \texttt{1}. Data is not modified in either of the two tests. The flags are only used to help the user filter out problematic entries at a later stage. 
}

{The \inline{check\_binary\_mismatch.txt} log file lists all the corrections that were done to infer the value of the \inline{binary} column, as well as the results of the tests performed.} The user can review the log file and force the value of \inline{binary} in  \inline{replacements.ini} if deemed necessary.

The next task is to query stellar catalogs to identify a common main identifier for each target star. 
To select a main identifier for the host star and the corresponding coordinates, \EMCv{1.0.0} first queried SIMBAD\footnote{\url{https://simbad.cds.unistra.fr/simbad/}}, then VizieR\footnote{\url{http://vizier.u-strasbg.fr/viz-bin/VizieR}} as a gateway to  the Kepler Input Catalog \citep{KIC}, the K2/EPIC Input Catalog \citep{K2},  and Gaia DR2 \citep{GAIA,GAIADR2}. \EMCv{1.0.0} queried these catalogs by the host star name of each entry; if the query was unsuccessful, it searched again by coordinates, using a ``ConeSearch''. The ConeSearch is a specific query that can be performed {that can be performed by all Virtual Observatory aware services}, which consists in searching by coordinates for the closest star inside a cone of a given tolerance radius. In \EMCv{1.0.0} the radius of the cone was gradually increased until all identifiers were found. This meant that, for the entries that did not have any counterpart available in the stellar catalogs, the script would associate a star that could have been far away from the planet candidate. Although this was the case for less than 100 identifiers in the cumulative catalog, it led to an erroneous result. 

In \EMCv{2.0.0}, we only consider SIMBAD and TIC v8.2 \citep{2021arXiv210804778P} as our stellar sources.
\EMCv{2.0.0}~first queries SIMBAD for the host name and the available aliases for each entry. The query also collects the official coordinates (epoch J2000) in degrees from SIMBAD.
This query is performed using all the available identifiers for a given target until a match is found, and including any information about binary stars if available. For all entries for which a match is found in SIMBAD searching by identifiers, we associate a value of \inline{angular\_separation} equal to \texttt{0.0} (since no coordinates were involved in the query) and we specify \texttt{SIMBAD} in the \inline{main\_id\_provenance} column.
The same query is repeated searching on the TESS Input Catalog (v8.2) for all entries whose SIMBAD search was unsuccessful. This catalog is the most recent input catalog available and it logs all the GAIA DR2 identifiers and coordinates, in addition to the TIC identifiers (TICid). If a match is found in TIC searching by TICid for a given entry, \inline{angular\_separation} is set to \texttt{0.0} and \inline{main\_id\_provenance} is set to to \texttt{TIC}.

{For the targets for which no SIMBAD nor TIC identifiers have been found, the script prepares for the query on the stellar catalogs based on coordinates. First, a coordinate check is performed (see above). If either the right ascension, the declination, or both coordinates do not match for all the entries reported to be orbiting the same host star, the flag \inline{coordinate\_mismatch\_flag} will be set to \texttt{1}, while the column \inline{coordinate\_mismatch} will store \texttt{RA}, \texttt{DEC}, or \texttt{RADEC} to inform on the specific coordinate(s) that failed the check. Entries that failed this check are also printed in the \inline{check\_coordinates.txt} file for user reference. No data is modified during this check. }

{A 1 arcsec radius ConeSearch on SIMBAD is then performed. The main identifier is selected to be the one with the smallest angular separation among all sources within the ConeSearch.} The separation values for all successful queries are saved in the  \inline{angular\_separation} column and  \inline{main\_id\_provenance} is set to \texttt{SIMBADCOORD}. 
Then, the same ConeSearch query is repeated on the TIC catalog for all the targets still missing a main identifier. Also in this case, if the 1 arcsec ConeSearch is performed successfully, the  \inline{angular\_separation} column is updated and the value of the column \inline{main\_id\_provenance} is set to \texttt{TICCOORD}. 
When identifiers are not found either through a name search or a ConeSearch in SIMBAD or TIC assuming a radius of 1 arcsec, \EMCv{2.0.0} does not increase the radius any further but assumes as main identifier and coordinates the ones stored in the input catalogs.  In this case the \inline{angular\_separation} column is set to \texttt{-1} and the \inline{main\_id\_provenance} is filled with a string labeling the input catalog from which the identifier is taken.

In \EMCv{2.0.0}, once the main identifier has been found, the script cleans the \inline{main\_id} values to ensure that no exoplanet identifiers (a small fraction of them also included in SIMBAD) are considered. The script also collects any extra information that the main identifier can provide, adding details on the binariety of the host star in the \inline{binary} column. This is automatically done if the current value of \inline{binary} is ``S-type'' or null. If the value in \inline{binary}  is non-null and disagrees with what found by the query on the stellar catalogs, this discrepancy is logged in the \inline{polish\_main\_id.txt} file, recommending user supervision. Then, the checks on the binary stars are repeated using the main identifier as a criterion for the grouping of the cumulative catalog (i.e. by grouping on \inline{main\_id}+\inline{letter}), to identify any further potential binary coordinate mismatches or complex systems.

Further checks are then performed to: 1) ensure that the same host star name is not found with different main identifiers, which would impair a correct merging of the entries (see Sect. \ref{sec:merging}); 2) ensure that the the same identifier is associated with the same coordinates, and 3) ensure that the same aliases are collected for the same main identifier. While these tests do not modify the catalog, they can be a useful diagnostic to the user in case issues occur in the stellar catalog query, or to identify peculiar cases. The results of these tests are logged in the \inline{post\_main\_id\_query\_checks.txt} file.

\subsection{Merging}\label{sec:merging}

What makes \EMC~useful is the merging of heterogeneous information from different source catalogs. For this reason, we have developed the merging algorithm that was used ever since the first release of the code. The goal of the merging is to obtain one single entry per target that collects the best estimate for each relevant parameter (mass, minimum mass, radius, period, semi-major axis, inclination, eccentricity) from the source catalogs and their references. The best estimate is the one that has the lowest relative error \citep[see][for details]{Alei2020}. In v2.0.0, we optimized this module to account for all the updates we implemented and to make the merging process more robust, by relying on physical parameters in addition to the ({homogenized}) names of each target.

Since we handle the exoplanet letter differently in \EMCv{2.0.0} (see Sect. \ref{sec:homogenization}), the same target could be present in the concatenated DataFrame with different values \inline{letter} (e.g. with a ``.0d'' string as well as with a lowercase letter). For this reason, \EMCv{2.0.0} first checks that the planet letter is the same for every entry having the same main identifier, binary string, and planetary period (or semi-major axis when no available period measurement is available). If only one lowercase letter appears for each group, any ``.0d'' string for the same group is replaced with the same lowercase letter. If there are more than one lowercase letter values that are inconsistent, these are not changed in the catalog, but are logged in the \inline{group\_by\_period\_check\_letter.txt} file and should be vetted by the user.

{In \EMCv{1.0.0}, the merging was executed based on the \inline{main\_id}, the \inline{binary}, and the \inline{letter} columns.  \EMCv{2.0.0} adds another layer of complexity to the merging to ensure that no mismatches are accidentally performed. The newly implemented merging routine checks that each group of entries that share 
\inline{main\_id}, \inline{binary}, and \inline{letter} also have the same period. If this is the case, the merging is performed. {If the period values are different more than 10\% of the value of each estimate (though the tolerance can be changed by the user), the merging is performed on the subgroups that share the same period value.} In the case where the period is not available, the grouping is performed using the semi-major axis. In the case where neither the period nor the semi-major axis are available, the merge is forced over all the elements of the group (hereafter: ``fallback merge''). }

All occurrences where there is a disagreement on period and/or semi-major axis, or when a fallback merge occurs, are logged in the \inline {group\_by\_letter\_check\_period.txt} file. {These occurrences are also recorded in two flags: the \inline{period\_mismatch\_flag} column, which is set to \texttt{1} when there has been a disagreement for the period or semi-major axis and multiple entries have been created; and the \inline{fallback\_merge\_flag}, set to \texttt{1} when no information on period or semi-major axis was available and the merge was forced.}

The merging module itself has undergone some minor updates. First of all, the naming convention changed: while in \EMCv{1.0.0} a name was picked as the most common one from the input catalogs,  \EMC~now creates the preferred target name by joining the main identifier, the binary value, and the letter value. The names by which every target was known in the original catalogs are also stored in their respective columns (see Table \ref{tab:catdesc}), a new addition to v2.0.0.{ Users can therefore query the input catalogs by name to gather further information or to retrieve the original dataset. 
}

The determination of the status was also updated, as two additional status flags (``controversial'' and ``preliminary'') are now available. \EMC~sets ``controversial'' status label for every entry that did not show agreement in the \inline{checked\_catalog\_status} (i.e., the updated status after the check on the KOI table). The label ``preliminary'' is instead reserved to elements in the source catalogs that have a null status. 
These are entries in the source tables that have been very recently added and are supposed to be updated in the near future. The original and checked catalog status values per each catalog are also collected in the \inline{original\_status\_string} and   \inline{checked\_status\_string} as a string representation of a dictionary (e.g., ``eu: confirmed, oec: confirmed''). This improves the readability and user-friendliness of the {output} catalog compared to \EMCv{1.0.0}, where the status was represented as a more cryptic \texttt{AXDXEXCX} string  \citep[``A'', ``D'',``E'',``C'' representing the four source catalogs and ``X'' being a number between 0 and 2 to represent the status of the planet in each source, see][for details]{Alei2020}. 

The \inline{discovery\_method} column (whose header changed for readability, see  Table \ref{tab:catdesc}) now allows for multiple detection methods to be listed, in case of independent detections.

When creating the single entry, the script checks that there are no duplicate values within the same input source. In case this happens, these are stored in the log file \inline{merge\_into\_single\_entry.txt}, the flag \inline{duplicate\_catalog\_flag} is set to \texttt{1} and the duplicate names are store in the \inline{duplicate\_names} column for easy filtering. This is mostly the case when both the candidate and the official planet name are stored in the catalogs. {Finally, the output catalog is checked to identify potential duplicate planets that have been not identified during the merging. This happens when there is missing information or discrepancies in the labeling of the binary star or the planet letter. The {output} catalog is grouped by \inline{main\_id}, then periods and semi-major axes are checked. If, in each group, similar values of \inline{p} or \inline{a} are found for entries that do not have agreeing values of \inline{binary} or \inline{letter}, the flag \inline{misnamed\_duplicates\_flag} is set to \texttt{1}. The output of this test is logged in \inline{identify\_misnamed\_duplicates.txt}. These cases should be vetted by the user and replacements should be listed in the \inline{replacements.ini} file to facilitate a correct merge. }

\subsection{Outputs}\label{sec:outputs}

In \EMCv{1.0.0}, we did not implement any cutoff on the mass and we inherently assumed the heterogeneous cutoffs of the input catalogs. 
In v2.0.0, we separate the brown dwarf candidates from the merged catalog based on a mass cutoff criterion of 20 $M_{Jup}$.

A single run of the \EMCv{2.0.0}~script produces three main files: first, the \inline{exo-mercat\_full.csv} file contains all the entries that \EMC~produced. Then, two ancillary files are produced: the \inline{exo-mercat\_brown\_dwarfs.csv} file contains all the entries whose (minimum) mass exceeded 20 $M_{Jup}$; the \inline{exo-mercat.csv} file contains all the other entries, including those that \EMC~identified as potential brown dwarfs (\inline{letter}=``BD'') but whose (minimum) mass was below the 20 $M_{Jup}$ cutoff. The user can filter the potential brown dwarfs out of the \inline{exo-mercat.csv} file by filtering the value of the \texttt{letter} column \emph{a posteriori}, or by forcing the removal of a specific target through the use of the \inline{replacements.ini} file (see Sect. \ref{sec:homogenization}).  By definition, the sum of the two ancillary files returns the full catalog.

 When an earlier {output catalog file produced by \EMC } is available, the script also compares row by row the current output catalog with the most recent available version and updates the value of the \inline{row\_update} column to the current date if any value in the row changed. This allows to keep track of the changes to every single row and to isolate the updated rows in every run.

The names of the column in {the catalog generated by} \EMC\ were updated in v2.0.0 to be more intuitive, and additional columns were added to the table. We explain the translation between the two versions in \ref{app:translation} (Table \ref{tab:catdesc}). 
As stated in the previous sections, the \EMCv{2.0.0}  script produces ten log files which inform the user whenever relevant changes on the data have been performed by the \EMC\  script, or whenever some flags have been raised. The names of the log files follow the ones of the functions that produced them. In Table \ref{tab:log_files} we summarize name and function of each log file. In addition to the log files, \EMCv{2.0.0} allows for on-screen logging using the \texttt{logging} package. This type of logging can be enabled when running the script with a command-line option \inline{--verbose} or \inline{-v}.

 \begin{table*} 
 \centering
 \begin{tabular}{p{0.3\textwidth}p{0.6\textwidth}p{0.05\textwidth} }
 \hline\hline {Log File} & {Content Summary} & {Sect.}\\ \hline 
 
 \inline{replace\_known\_mistakes.txt} & \hangindent=1em  Lists unused replacements during the {homogenization} of the sources. & \ref{sec:homogenization}\\ 
 \inline{alias\_as\_host.txt} & \hangindent=1em  Lists host stars found by other names in input catalogs. & \ref{sec:mainid}\\ 
  \inline{check\_binary\_mismatch.txt} & \hangindent=1em  Lists corrections for the \inline{binary} column and warnings for coordinate discrepancies.  & \ref{sec:mainid}\\ 
 \inline{check\_coordinates.txt} & \hangindent=1em  Contains entries that failed coordinate checks before SIMBAD and TIC queries. & \ref{sec:mainid} \\
 \inline{polish\_main\_id.txt} & \hangindent=1em  Lists entries whose \inline{main\_id} was cleaned and where new \inline{binary} values were identified.  & \ref{sec:mainid}\\ 
 \inline{post\_main\_id\_query\_checks.txt} & \hangindent=1em Lists checks performed after finding main identifier, including duplicate hosts and similar coordinates.  & \ref{sec:mainid} \\ 
\inline{group\_by\_period\_check\_letter.txt} & \hangindent=1em Logs systems with inconsistencies in planet letter for the same period.  & \ref{sec:merging}\\
 \inline{group\_by\_letter\_check\_period.txt} & \hangindent=1em  Contains groups with disagreements in period or semi-major axis. & \ref{sec:merging}\\
 \inline{merge\_into\_single\_entry.txt} & \hangindent=1em  Lists duplicate entries within the same input catalog. & \ref{sec:merging}\\
  \inline{identify\_misnamed\_duplicates.txt} & \hangindent=1em  Lists potential missed duplicates in \EMC~because of mismatches in the planet name (\inline{binary} and \inline{letter}). & \ref{sec:merging}\\\hline
 \end{tabular} 
 \caption{Summary of the log files produced \EMCv{2.0.0} and their function.} \label{tab:log_files} \end{table*}

\subsection{Folder structure}

In this new version, the user can impact the performance of \EMC~according to the specific needs. It is also easy to add custom sources to the calculation, if desired.

The script was refactored into modules and classes, included in the main module folder \inline{exomercat}. Documentation and typesetting hints were also added, to increase readability.
The six input catalogs, as well as the {output catalog generated by \EMC}, are now subclasses of a parent class named \inline{Catalog}. The \inline{Catalog} class contains all the common methods for the currently implemented catalogs. These will be read when {homogenizing} the specific input source unless overwritten by one other subclass, specific to that input source. The subclasses (one per catalog and in different python files) describe all the catalog-specific methods. These methods will raise a \inline{NotImplementedError} if not implemented in the subclass. 
This way of refactoring the script paves the way for more input sources to be added in the future. The \inline{NotImplementedError} can be used to point out all the methods that need to be specified when developing the ingestion of a new catalog source.

The utility functions are collected as static methods within the \inline{UtilityFunctions} class. These are service functions that are used by one or more classes to perform specific tasks (e.g. read a \texttt{.ini} file, gather information from \texttt{.xml} files).

{The command-line functions that can be used to run \EMC~are stored in the \inline{cli.py} file. These include: 1) the \inline{ping} function, used to run checks on the current status of the input sources and the stellar catalogs; 2) the \inline{input} function, which downloads and {homogenizes} the tables (see Sects. \ref{sec:ingestion} and \ref{sec:homogenization}); the \inline{run} function, which loads {homogenized} tables and creates the {output} catalog (see Sects. \ref{sec:mainid} to \ref{sec:outputs}); 4) the \inline{check} function, which runs general sanity checks on the output catalog.}

During runtime, the most up-to-date versions of the source catalogs are downloaded and stored in the \inline{InputSources} folder. The source catalogs that were successfully refined to a common output {layout} are stored in the \inline{StandardizedSources} folder. For each run, a set of log files is produced in the \inline{Logs} folder.  At the end of every run, the catalog outputs are stored in the \inline{Exo-MerCat} folder. When saving, the input sources and the and merged catalog file names contain also the date of the script execution in \texttt{YYYY-MM-DD} format. 
 
The released version has been thoroughly tested with unit tests, which are stored in the \inline{tests} folder.

The \texttt{.ini} configuration files used to configure the download of the input sources (\inline{input\_sources.ini}) and to force known replacements to be performed (\inline{replacements.ini}) are by default in the root folder, together with a README file that lists the major changes in each version. {If \EMC~is installed as a package, these files could be moved to any location\footnote{For details, see: \url{https://exo-mercat.readthedocs.io/}}.}

% \dirtree{%
% .1 Exo-MerCat.
% .2 exo\_mercat.
% .3 catalogs.py.
% .3 emc.py.
% .3 epic.py.
% .3 eu.py.
% .3 koi.py.
% .3 nasa.py.
% .3 oec.py.
% .3 toi.py.
% .3 utility\_functions.py.
% .2 Exo-MerCat.
% .3 exo-mercatMM-DD-YY.csv.
% .3 $...$ .
% .2 InputSources.
% .3 nasa\_initMM-DD-YY.csv.
% .3 $...$ .
% .2 StandardizedSources.
% .3 nasa.csv.
% .3 $...$ .
% .2 Logs.
% .3 replace\_known\_mistakes.txt.
% .3 $...$ .
% .2 tests.
% .2 input\_sources.ini.
% .2 main.py.
% .2 replacements.ini.
% .2 README.md.
% }

 % We report the translation table between the two versions in Table \ref{tab:catdesc}. Here, we indicate with a star symbol ($\star$) all of the columns that changed their meaning between the two versions of the script. 

\section{Discussion}\label{sec:discussion}
\subsection{Comparison to previous version}\label{sec:comparison}

While the philosophy behind \EMC~has not changed, the \EMCv{2.0.0} script produces a catalog that contains more data (because of the inclusion of more input sources) and a more complete set of measurements for each target, compared to the previous versions. In the following, we compare the catalog generated by \EMCv{2.0.0}  with the one generated by the previous version of the script, which has been updated in order to: 1) exclude the retired Exoplanet Orbit Database catalog; and 2) replace the URLs to the NASA, EU, and KOI catalogs to make the script not fail at runtime. We refer to this specific older version as \EMCv{1.1.0}. The {catalog generated by} \EMCv{2.0.0} that we use for this comparison is {contained in} \inline{exo-mercat_full.csv} {and it} includes also the brown dwarfs, for a fairer comparison with the {catalog generated by }\EMCv{1.1.0}. The \EMCv{2.0.0} and \EMCv{1.1.0} scripts were run on the same date (October 25, 2024), hence using the same input sources tables (though not all tables were used by \EMCv{1.0.0}, see Sect. \ref{sec:ingestion}). Details on the \EMCv{2.0.0} run as of October 25, 2024 can be found in  \ref{sec:rundetails}.

In Table \ref{tab:comparison} we compare the five input catalogs with {the ones generated by}  \EMCv{1.1.0} and \EMCv{2.0.0}. The {catalog produced with the latest version} shows an increase in the amount of planets compared to \EMCv{1.1.0} of nearly a factor two, mostly thanks to the addition of the TOI input source. The number of confirmed planets increases of about 250 targets, while the number of candidate and false positive targets increases of approximately a factor three or more than ten respectively. The entries that have an estimate of the radius or planetary period is also doubled, while the ones that possess a mass or minimum mass estimate are, in number, very similar for the two versions. Again, this variation is connected to the addition of the TOI catalog, which provides periods and radii for most of the targets that it contains. 

\begin{figure}
    \centering
    \includegraphics[width=1\linewidth]{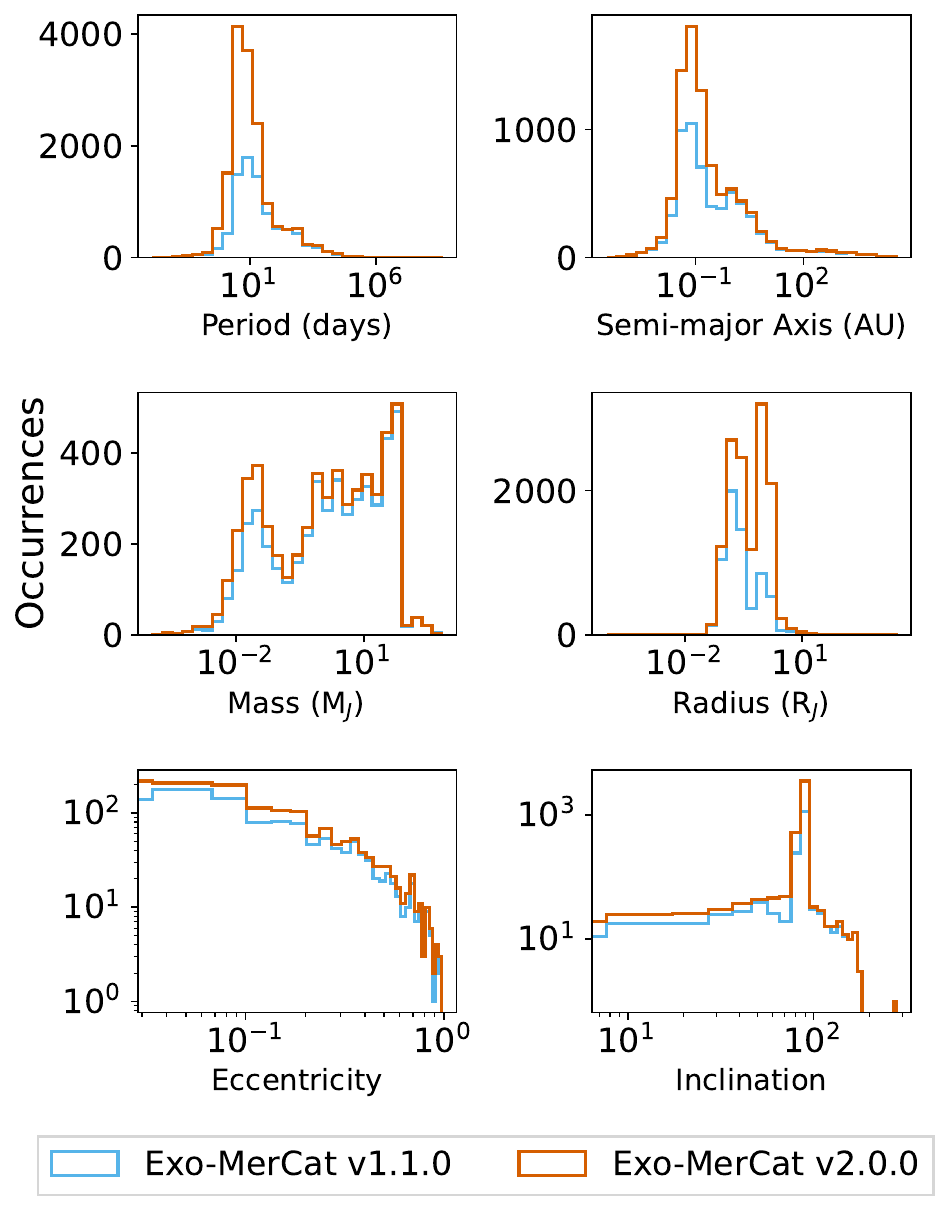}
    \caption{Histogram distributions for the period (top left), semi-major axis (top right), mass (center left), radius (center right), eccentricity (bottom left), and inclination (bottom right) for {the catalogs generated by} \EMCv{1.1.0} (in blue) and \EMCv{2.0.0} (in red). Dataset updated on October 25, 2024. }
    \label{fig:multiplot}
\end{figure}

\begin{figure}
    \centering
    \includegraphics[width=1\linewidth]{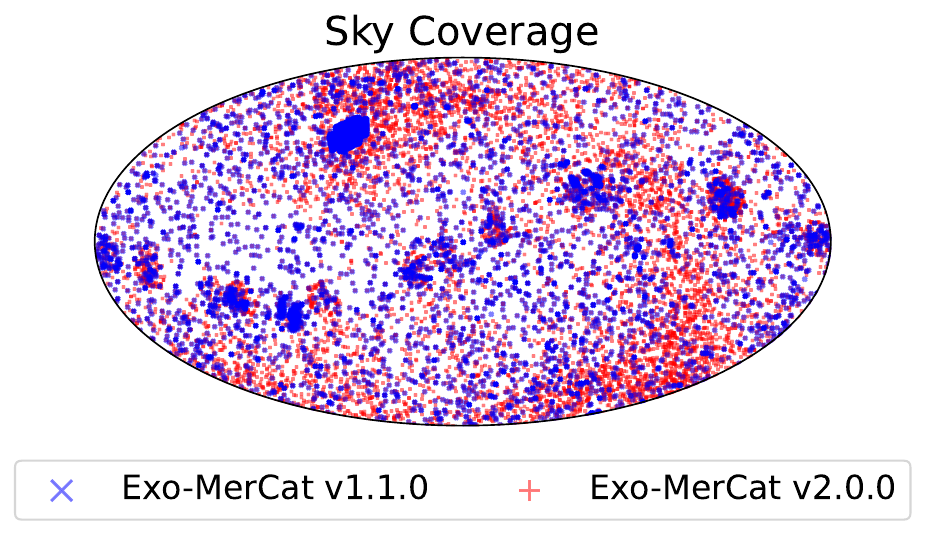}
    \caption{Sky Coverage comparison between {the catalogs generated by }\EMCv{1.1.0}~(in blue) and \EMCv{2.0.0}~(in red). Dataset updated on October 25, 2024.}
    \label{fig:versioncomparison}
\end{figure}

In Figure \ref{fig:multiplot} we compare the distribution of the main planetary parameters between the {output catalogs produced by the} two versions of \EMC. In this visualization, it is clearly visible the added benefit that \EMCv{2.0.0} (red histogram) has on the completeness of the sample, especially with respect to the period, semi-major axis, and radius. In Figure \ref{fig:versioncomparison} we compare the distribution on sky of {the catalogs generated by} the two \EMC~scripts (\EMCv{1.1.0} in blue, \EMCv{2.0.0} in red). Here, the larger number of targets included in {the catalog generated by} \EMCv{2.0.0} and their position on sky is immediately noticeable and follows the distribution of the TOI candidates known to date.

\begin{table*}
    \centering
    \begin{tabular}{lccccccc}\hline\hline
Query & NASA & EU & OEC & TOI & EPIC & v1.1.0 & v2.0.0\\\hline
All planets & 5785 & 9806 & 5402 & 7241 & 1800 & 9900 & 17445 \\
Confirmed & 5785 & 7373 & 5288 & 1012 & 575 & 7383 & 7621 \\
Candidates & 0 & 2329 & 100 & 4621 & 975 & 2402 & 7632 \\
False Positives & 0 & 104 & 14 & 1176 & 250 & 115 & 1488 \\
With semi-major axis & 5499 & 7083 & 2802 & 0 & 390 & 6113 & 8687 \\
With radius & 4379 & 6551 & 4155 & 6771 & 1434 & 6553 & 13435 \\
With mass & 1999 & 4811 & 2505 & 0 & 130 & 4779 & 5449 \\
With msin(i) & 852 & 1632 & 261 & 0 & 14 & 1671 & 1717 \\
With period & 5504 & 8222 & 5095 & 7141 & 1777 & 8317 & 15702 \\
With mass or msin(i) & 2851 & 4811 & 2766 & 0 & 138 & 4895 & 5489 \\
With mass and msin(i) & 0 & 1632 & 0 & 0 & 6 & 1555 & 1677 \\
With mass and msin(i) and radius & 0 & 16 & 0 & 0 & 0 & 18 & 0 \\
With mass or msin(i) and radius & 1469 & 1798 & 1544 & 0 & 127 & 1847 & 2487 \\
With mass or msin(i), and radius and period& 1451 & 1461 & 1469 & 0 & 127 & 1519 & 2105 \\
All systems & 4319 & 7717 & 4059 & 6963 & 1560 & 7614 & 14483 \\\hline
    \end{tabular}
    \caption{Available measurements for various combinations of parameters in the five catalogs as they were downloaded from their sources, plus {the catalogs produced by \EMCv{1.1.0} and \EMCv{2.0.0}}. Dataset updated on October 25, 2024.}
    \label{tab:comparison}
\end{table*}

% \begin{figure*}
%     \centering
%     \includegraphics[width=1\linewidth]{Emc2_sky_coverage.pdf}
%     \caption{Enter Caption}
%     \label{fig:inputcatalogcomparison}
% \end{figure*}

\begin{figure}
    \centering
    \includegraphics[width=1\linewidth]{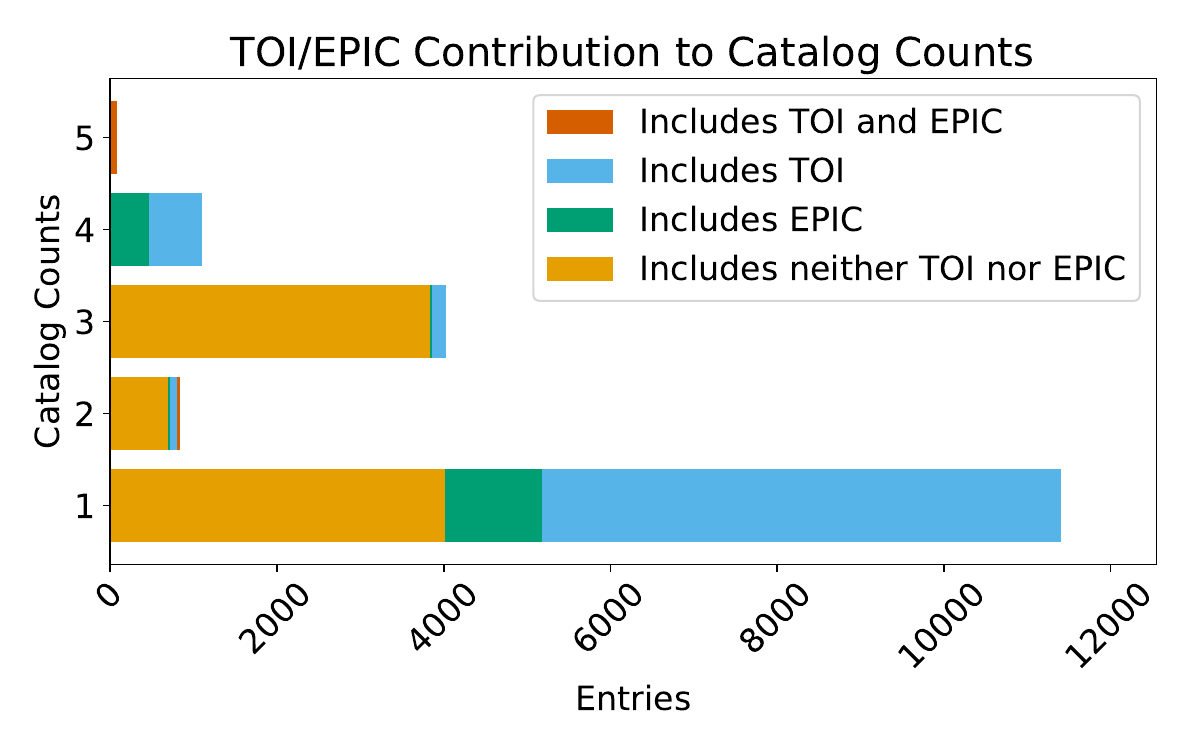}
    \caption{TOI and EPIC contribution to the total sample, measured in catalog counts (from one to five occurrences total in the source catalogs). In red, the sample that includes both TOI and EPIC (potentially together with other input sources); in blue, the sample that includes TOI (potentially together with other input sources) but not EPIC; in green, the sample that includes EPIC (potentially together with other input sources) but not TOI; in orange, the sample that does not include TOI nor EPIC. Dataset updated on October 25, 2024.}
    \label{fig:catalog_count}
\end{figure}

In Figure \ref{fig:catalog_count} we show a stacked bar histogram that highlights the distribution of entries in {the catalog generated by} \EMCv{2.0.0} depending on the number of input catalogs where these were found (e.g., merged entries that were found in all the input catalogs have a value of 5; entries that were found in only one input catalog have a value of 1). In different colors, we highlight the contribution of the TOI and EPIC catalogs in the final distribution. In orange, we show all the entries that did not have a counterpart in TOI nor EPIC, but were only found in one or more of the NASA, EU, and OEC catalogs. Because of this, we can only see entries in the orange subset that have a value smaller or equal to 3. In blue, we highlighted the targets that were found in TOI but not in EPIC, while we represent the targets found in EPIC but not in TOI in green. The targets in the green and blue sample can be found also in NASA, EU, and OEC (catalog count is 4), though the majority are found only in TOI or EPIC respectively (catalog count is 1). 
A small portion of targets is found in both EPIC and TOI, shown in red  in Figure \ref{fig:catalog_count}. This subset is divided into 80 targets that have been found in all catalogs (catalog count is 5) and 38 that have been found only in TOI and EPIC (catalog count is 2).
We observe that the contribution of TOI and EPIC catalog does not overlap substantially with the other catalogs, since most of the entries appear in the single-catalog bar (catalog count is 1). However, there is a fraction of candidates that are found in NASA, EU, and OEC, as well as one or both TOI and EPIC (catalog counts is 4 or 5). In this case, the information included in the two additional catalogs has enriched the content of the existing entries, which is one of the fundamental differences between \EMCv{1.1.0} and \EMCv{2.0.0}.

%{Other major updates in \EMCv{2.0.0} were the optimization of the merging (see Sect. \ref{sec:merging}) and the addition of more descriptive flags (see Table \ref{tab:comparison} for details) which can be used to filter out troublesome entries. }%This allowed us to compare a few candidates that were identified as duplicated entries in \EMCv{2.0.0} (i.e., which had \inline{duplicate\_catalog\_flag}=1) but were identifies as separated entries in \EMCv{1.0.0}. Some example are:

% \begin{itemize}
%     \item \inline{exo-mercat_name}=* iot Hor b.  In \EMCv{2.0.0}, this planet is found to have a duplicate in the EU catalog under ``HR 810 b'' and ``Gaia-ASOI-004 b''. In \EMCv{1.0.0}, two separate entries appear, the first as ``Gaia-ASOI-0004 b'' and the second containing ``HR 810 b''. The identification of the duplicate was possible through the association of the period.
% \end{itemize}

\subsection{Value to the community}\label{sec:value}

Since the release of v1.0.0, \EMC~has been used in a few publications. For example, \citet{2023A&A...675A.158D}, \citet{2021A&A...650A..66B}, and \citet{2021A&A...645A..71C} used the catalog {generated by \EMC} to identify a subset of known transiting planets that satisfied specific conditions on mass, radius, and eccentricity to put transiting planets detected around young stars in context. \citet{app11083322} and \citet{2020AJ....160..246M} used \EMC~to quantify the amount of confirmed and candidate planets known to date. 

\EMC~is also used by the PLAnetary Transits and Oscillations of stars (PLATO) and the Large Interferometer For Exoplanets (LIFE) science teams as source to their Input Catalogs. 

PLATO is an adopted ESA mission planned for launch in 2026, focused on detecting and characterizing transiting exoplanets in the habitable zone of Sun-like stars \citep[see e.g.,][for details]{2024arXiv240605447R}. In the PLATO Input Catalog (PIC), \EMC~is used to identify all the stars known to host confirmed or candidate planets (including non-transiting ones) with the aim of including them for observations, even if their properties are not compliant with the criteria assumed for the PLATO scientific targets \citep[see][]{2021A&A...653A..98M}. Though {the default} \EMC{-generated catalog} does  not contain information on the stellar host, it is essential to have a list of host star identifiers to query in the stellar catalogs. At the time of writing, the PIC developers have already upgraded the \EMC~version to v2.0.0 and can benefit from the higher quality merge performed by \EMCv{2.0.0}~which allows to unambiguously identify each target with their original names and status identifiers in the exoplanet source catalogs. All of this allows them to gather additional information on the star from the exoplanet source catalogs and stellar catalogs through automated scripts and requiring little to no human supervision. The  \inline{duplicate\_catalog\_flag}, \inline{duplicate\_names}, \inline{period\_mismatch\_flag} and \inline{misnamed\_duplicates\_flag} columns allow them to make informed choices on potentially duplicated data or missed merges. {In addition, \EMC~was used to perform statistics on known and candidate planets in the LOPS2, i.e. the first Long Pointing PLATO field (see {Nascimbeni et al. 2024, submitted}).}

LIFE is a concept for a space-based nulling interferometer, aiming at detecting and characterizing the thermal emission of terrestrial planets around F, G, K, and M stars in direct imaging \citep[see e.g.,][for details]{2022A&A...664A..21Q}. The LIFE Target Database is being developed to provide a list of preliminary potential targets for the LIFE mission. Its goal is to merge information for stars, disks, and planets, that would be part of the LIFE Target List. \EMC~is used to select all the stars known to host planets. It is especially useful to the team since the most challenging part, the comparison and merging of the data among various exoplanet catalog, has already been performed and {homogenized}. In future releases of the LIFE Target Database, \EMC~will be used to prioritize targets for the LIFE observations, e.g., higher priority will be given to stars where interesting exoplanets are detected and lower priority to those that are known to host Hot Jupiters and where theory predicts no habitable terrestrial planets should be present.

{The capability of \EMC~of merging together different exoplanet catalogs, by {homogenizing} them, is also useful for webtools whose main aim is to search for exoplanetary system architectures.
Among these we can cite the ASI-SSDC ExoplAn3T (Exoplanetary Analysis and 3D Tool\footnote{\url{https://tools.ssdc.asi.it/exoplanet}}) which, by querying remote archives, such as the NASA and EU catalogs, together with \EMC, is able to provide lists of exoplanetary systems based on user-defined planetary, stellar, or orbital characteristics.
ExoplAn3T has already demonstrated its usefulness for the study of the dynamical history of a series of exosystems (e.g., \cite{Zinzi2017}, \cite{Turrini2020}, \cite{Turrini2023}) and can also provide hints in the field of astrobiology studies, when used to look for exosystems with architectures similar to the Solar System, or that host potentially habitable planets \cite{Zinzi2021}.}

{The \EMC~script can benefit a variety of astronomers: some might find the catalogs that are maintained and regularly updated to the TAP by the \EMC~team useful; others might instead prefer to download the script, to execute it locally and make changes. In this case, quality control of the results is also passed down to the user.}
\subsection{Code support and future improvements}

\label{sec:opensource}

We released \EMCv{2.0.0}~on GitHub\footnote{\url{https://github.com/Exo-MerCat}} as open source, together with other useful modules \citep[e.g., the Graphic User Interface, see][for details]{Alei2020}. Users are encouraged to fork the repository and work with the code. Any update on the existing GitHub will be reviewed through the ``merge request'' process. This includes updates on the \inline{replacements.ini} file, if further replacements are found when running the code.

The code {runs with Python 3.8 and above }and it is fully tested through unit tests. These are all expected to pass for a stable version of the script. In the case of an update of any input source, an \EMC~run might still fail. The user can report the issue in the ``Issues'' panel.

The code has been fully documented. The main documentation can be found in the ReadTheDocs webpage\footnote{\url{https://exo-mercat.readthedocs.io/en/latest/}}.

Both {catalogs produced by} \EMCv{1.1.0} and \EMCv{2.0.0} are reachable through a TAP service, hosted at IA2 Data Center\footnote{\url{https://www.ia2.inaf.it/}}, in a similar way as described in \citep{Alei2020}.
The TAP service is the same for both versions\footnote{\url{http://archives.ia2.inaf.it/vo/tap/projects}}
and the two versions are available as different resources. {The catalog generated by }\EMCv{1.1.0}~is stored as \texttt{exomercat.exomercat\_legacy} and is no longer actively maintained, but users can still query the latest available update. \EMCv{2.0.0} {is executed monthly and two tables are uploaded on the TAP:} \texttt{exomercat.exomercat}, which contains the result of the latest run, i.e. the most updated snapshot of the \EMC{-generated} catalog; and \texttt{exomercat.exomercat\_global}, which collects all the entries that were updated on each run, without repeating identical entries from different runs. The structure of the two new tables is the same. Compared to the columns in the output \texttt{.csv} file of \EMC, both tables will feature an additional column, (\texttt{exomercat\_version}) to log the version of the code that produced the result. This way, the user can easily retrieve the history of all modifications that a single target underwent in time. We plan also to store separately all the monthly runs of \EMCv{2.0.0}. This will give us the possibility in the future to create and implement a service to allow users to download a snapshot of specific time in the past.

Future improvements on the code include: the addition of ancillary sources; the production of log files that are specific to the various input sources, which can be used to inform the maintainers of input sources of errors in the input catalogs in a more streamlined way; and the addition of further user cases and tutorials in the documentation.
\EMC~will be indexed in all major astronomic software databases such as the Exoplanet Modeling and Analysis Center \citep[EMAC,][]{2022RNAAS...6..185R} and the Astrophysics Source Code Library \citep[ASCL,][]{1999AAS...194.4408N,2020ASPC..522..731A} for a wider spread in the community. 

\section{Summary}\label{sec:conclusion}

In this work, we introduced the newest version of \EMC, {a script} that collects data from the most known exoplanet databases and {homogenizes} the sample in a VO-aware way. The most relevant upgrades that differentiate \EMCv{2.0.0} from the previous versions of the code are summarized here: 
\begin{itemize}
    \item The TESS Project Candidates Table was added to the input sources, and the EPIC/K2 Planets and Candidates Table was promoted from look-up table to input source. The Exoplanet Orbit Database was removed from the input sources, since it has since been retired. The interface with NASA Exoplanet Archive, the Kepler Objects of Interest catalog, and the Exoplanet Encyclopaedia was also updated to comply with the changes in those databases. This upgrade increases the number of entries in the {output} catalog and the overall completeness of the parameter estimates in the sample.
        \item The main identifier is now searched in SIMBAD and TESS Input Catalog v8.2 for both name- and coordinate-based queries. In the case of coordinate-based queries, the query is performed on a much smaller tolerance radius (1 arcsec instead of 36 arcsec). This reduces the amount of mistakes in the attribution of the main identifier compared to the previous version.
    \item The grouping of the entries now takes into account the estimates of the period or semi-major axis. This ensures a more accurate merge that overcomes discrepancies in the planet letter.
    \item \EMC~now produces a filtered version of the catalogs that removes all brown dwarf candidates with mass smaller than 20 $M_{Jup}$, in addition to the complete catalog. Extra flags have been added to the final table and allow for custom filtering.

    \item The script produces more user-friendly log files, which can be used to force corrections to enable the correct merging of the entries. The user can easily handle these corrections through ancillary files. This update adds more flexibility and user-friendliness to the script.
    \item The code has been restructured into modules and functions. This upgrade enables the inclusion of more input sources in the future.
    \item \EMCv{2.0.0}~has been released open-source and it is currently fully tested and documented.

\end{itemize}

\EMC~has been used for various studies since the first publication. The catalog is also a source for the PLATO Input Catalog and the LIFE Target Database. The release of the new version of the code would be a valuable resource for other mission input catalogs, such as the Habitable Worlds Observatory Preliminary Input Catalog \citep[HPIC,][]{Tuchow_2024}. The open-source release of \EMC~also allows for flexibility in the long-term maintenance of the code, which can be fully customized to the needs of every team.

\section*{Acknowledgments}
E. Alei thanks Sascha P. Quanz and Hans Martin Schmid for helpful conversations. E. Alei’s work has been partly carried out within the framework of the NCCR PlanetS supported by the Swiss National Science Foundation under grants 51NF40\_182901 and 51NF40\_205606.  E. Alei’s research was partly supported by an appointment to the NASA Postdoctoral Program at the NASA Goddard Space Flight Center, administered by Oak Ridge Associated Universities under contract with NASA.
S. Marinoni acknowledges financial support from the ASI-INAF agreement n. 2022-14-HH.0.

During the preparation of this work the authors used ChatGPT in order to produce code documentation, refactor code, and produce unit tests. After using this tool/service, the authors reviewed and edited the content as needed and take full responsibility for the content of the publication.

\emph{CRediT roles:} E. Alei: Conceptualization, Data curation, Formal Analysis, Investigation, Methodology, Project administration, Software, Validation, Visualization, Writing - original draft, Writing - review and editing; S. Marinoni: Investigation, Methodology, Validation, Writing - original draft, Writing - review and editing; A. Bignamini: Data curation, Funding acquisition, Project administration, Resources, Software, Writing - original draft, Writing - review and editing; R. Claudi: Conceptualization, Funding acquisition, Project administration, Supervision, Writing - review and editing; M. Molinaro: Data curation, Resources, Software, Writing - review and editing; M. Vicinanza: Data curation, Software; S. Benatti: Writing - review and editing; I. Carleo:  Writing - review and editing; A. Mandell:  Writing - review and editing; F. Menti:  Writing - review and editing; A. Zinzi:  Writing - review and editing.
\appendix

\section{Column headers translation} \label{app:translation}

In Table \ref{tab:catdesc} we list the previous and current naming convention for the {catalog generated by} \EMC. We indicate with a star symbol ($\star$) all of the columns that changed their meaning between the two versions of the script. We refer to the main text for more details on these changes.

\section{Log for the October 25, 2024 run}\label{sec:rundetails}

As of October 25, 2024, the concatenated DataFrame contained 30034 entries, which were thus distributed: 9804 entries from EU, 7241 entries from TOI,  5785 entries from NASA,  5402 entries from OEC, 1800 entries from EPIC.
Out of 15479 unique host star names, 2088 entries had as host star name one of the aliases in other entries and were set to the same value.

A total of 649 systems with an inconsistent value of ``S-type'' or ``null'' binary together with non-null values (641 occurrences) or only ``S-type'' and ``null'' occurrences (8 occurrences) could be fixed automatically. Of these, 370 systems (1085 single entries) showed coordinates that were outside the expected tolerance (\texttt{binary\_coordinate\_mismatch\_flag}=1). This check identified also 19 complex systems (70 total entries with \texttt{binary\_complex\_system\_flag}=1). 

One entry (2MASS J0441+2301 A b) failed the coordinate check before the main identifier search for both right ascension and declination (with a separation of about 11 arcsec between the occurrences). 

The main identifier search showed the following results: 28158 out of 30034 of the identifiers in the cumulative DataFrame were found by querying SIMBAD by host name or alias (94\% sample completeness); 1524 more were found by querying TIC by host name or alias (98.9\% sample completeness), 163 more were found via coordinates (99.4\% sample completeness). Only 189 entries out of 30034 in the concatenated table could not be found in the stellar catalogs. For these, the identifiers were taken from the input catalogs.

The binary mismatch check on the main identifier allowed the fix of 19 more S-type systems, of which 13 had differences in the coordinates (\inline{binary\_coordinate\_mismatch\_flag}=1); 38 additional complex systems were found (\inline{binary\_complex\_system\_flag}=1). The main identifier suggested the presence of 28 potential binary stars that were not considered in the original sources.
Six pairs of stars were found to have the same value of \inline{host} but different values of \inline{main\_id}, either complex systems or stars are discovered to be gravitationally bound only after their discovery.

The planet letters were incoherent in 1094 occurrences. In 1020 of these, the planets in the cumulative catalog were labeled with their candidate string (``.0d'') as well as their confirmed string (a lowercase letter between ``b'' and ``j''). In these cases, \EMC~forced the value of letter to be the one of the confirmed planet. The 74 remaining cases could not be automatically {homogenized} and were logged to be fixed manually.

The final table had 17745 entries, whose status is thus distributed: 7621 confirmed planets; 7632 candidate planets; 1488 false positives; 702 controversial planets; and two preliminary planets.  When merging the entries to create the {final} catalog, 197 groups disagreed on the period or semi-major axis (\inline{period_mismatch_flag}=1) and resulted in 399 separate entries; 16 groups had a fallback merge. In total, these entries constituted about 2\% of the final number of entries in {the catalog generated by} \EMC. A total of 62 entries had duplicate values in one or more input catalogs. 


%% If you h
 
\clearpage
\onecolumn
\begin{longtable}{l|l|p{8cm}|l}
\caption{Column headers {used by }\EMCv{1.0.0}, \EMCv{2.0.0}, their meaning, and their type.\label{tab:catdesc}}\\
  \hline\hline
  \EMCv{1.0.0}  & \EMCv{2.0.0} & Meaning & Type\\\hline
\inline{name}& \inline{exo-mercat\_name}& Planet name chosen by \EMC & \inline{str}\\
- & \inline{nasa\_name}& Planet name in the NASA Exoplanet Archive & \inline{str} \\
- & \inline{toi\_name}& Planet name in the TESS Objects of Interest Table & \inline{str} \\
- & \inline{epic\_name}& Planet name in the Kepler/K2 Objects of Interest & \inline{str} \\
- & \inline{eu\_name}& Planet name in the Exoplanet Encyclopaedia & \inline{str} \\
- & \inline{oec\_name}& Planet name in the Open Exoplanet Catalog & \inline{str} \\
\inline{host}& \inline{host} & Host star name& \inline{str}\\
\inline{letter}& \inline{letter}& Letter labeling the planet& \inline{str}\\
\inline{main\_id}  & \inline{main\_id}  & $\star$  Main identifier of the host star from SIMBAD or TIC catalogs & \inline{str}\\
 \inline{binary}& \inline{binary} & String labeling the binary host star, if any& \inline{str}\\
\inline{ra\_off}& \inline{main\_id\_ra} & $\star$ J2000 right ascension in degrees from SIMBAD & \inline{float}\\
\inline{dec\_off}& \inline{main\_id\_dec} & $\star$ J2000 declination in degrees from SIMBAD& \inline{float}\\
\inline{mass}& \inline{mass}& Planet mass in Jovian masses& \inline{float}\\
\inline{mass\_max}& \inline{mass\_max}& Positive error on the mass in Jovian masses& \inline{float}\\
\inline{mass\_min}&\inline{mass\_min}& Negative error on the mass in Jovian masses& \inline{float}\\
\inline{mass\_url}&\inline{mass\_url}& Bibcode of the reference paper for the mass value& \inline{str}\\
\inline{msini}&\inline{msini}& Planet minimum mass in Jovian masses& \inline{float}\\
\inline{msini\_max}& \inline{msini\_max}& Positive error on the minimum mass in Jovian masses & \inline{float}\\
\inline{msini\_min}& \inline{msini\_min} & Negative error on the minimum mass in Jovian masses& \inline{float}\\
\inline{msini\_url}& \inline{msini\_url} & Bibcode of the reference paper for the minimum mass value& \inline{str}\\
\inline{bestmass}& \inline{bestmass} & Most precise value between mass and minimum mass in Jovian masses& \inline{float}\\
\inline{bestmass\_max}& \inline{bestmass\_max}& Positive error on the best mass in Jovian masses& \inline{float}\\
\inline{bestmass\_min}& \inline{bestmass\_min}&  Negative error on the best mass in Jovian masses& \inline{float}\\
\inline{bestmass\_url}& \inline{bestmass\_url}&Bibcode of the reference paper for the best mass value& \inline{str}\\
\inline{mass\_prov}&  \inline{bestmass\_provenance}& String labeling the origin of the best mass (mass or minimum mass)& \inline{str}\\
\inline{p}&\inline{p} & Planet orbital period in days& \inline{float}\\
 \inline{p\_max}& \inline{p\_max}&Positive error on the period in days& \inline{float}\\
 \inline{p\_min}& \inline{p\_min} & Negative error on the period in days& \inline{float}\\
 \inline{p\_url}& \inline{p\_url} & Bibcode of the reference paper for the period value & \inline{str}\\
 \inline{r}& \inline{r} & Planet radius in Jovian radii& \inline{float}\\
 \inline{r\_max}& \inline{r\_max}& Positive error on the radius in Jovian radii& \inline{float}\\
 \inline{r\_min}&\inline{r\_min}& Negative error on the radius in Jovian radii & \inline{float}\\
 \inline{r\_url}&\inline{r\_url}& Bibcode of the reference paper for the radius value& \inline{str}\\
 \inline{a}& \inline{a}& Planet semi-major axis in au & \inline{float}\\
 \inline{a\_max}& \inline{a\_max}& Positive error on the semi-major axis in au& \inline{float}\\
 \inline{a\_min}& \inline{a\_min}& Negative error on the semi-major axis in au& \inline{float}\\
 \inline{a\_url}&\inline{a\_url} & Bibcode of the reference paper for the semi-major axis value & \inline{str}\\
 \inline{e}& \inline{e}& Eccentricity of the planet (scalar between 0 and 1)& \inline{float}\\
 \inline{e\_max}& \inline{e\_max}&  Positive error on the eccentricity (scalar)& \inline{float}\\
 \inline{e\_min}& \inline{e\_min} & Negative error on the eccentricity (scalar)& \inline{float}\\
 \inline{e\_url}& \inline{e\_url}& Bibcode of the reference paper for the eccentricity value & \inline{str}\\
 \inline{i}& \inline{i}& Planet inclination in degrees.& \inline{float}\\
 \inline{i\_max}& \inline{i\_max}&  Positive error on the inclination in degrees & \inline{float}\\
\inline{i\_min}& \inline{i\_min} & Negative error on the minimum mass in degrees& \inline{float}\\
 \inline{i\_url}&\inline{i\_url}& Bibcode of the reference paper for the inclination value & \inline{str}\\
\inline{discovery\_method}& \inline{discovery\_method} & Planet discovery method& \inline{str}\\
\inline{Status}& \inline{status}  & $\star$ Planet status of the planet preferred by \EMC& \inline{str}\\
 \inline{Status\_string}& \inline{original\_status\_string} & $\star$ String listing the planet status in all available input sources & \inline{str} \\
- & \inline{checked\_status\_string} & $\star$ String listing the planet status in all available input sources after the KOI check & \inline{str} \\
 
\inline{confirmed}& \inline{confirmed} & $\star$ Number of CONFIRMED status in \inline{checked\_status\_string} & \inline{int}\\
 \inline{yod}& \inline{discovery\_year}& Planet discovery year& \inline{int}\\
\inline{alias}& \inline{main\_id\_aliases} & String listing all known aliases for the host star& \inline{str} \\
 - & \inline{main\_id\_provenance} & Provenance of the main identifier & \inline{str} \\
  - & \inline{angular\_separation\_flag} & Flag for non-null values of \inline{angular\_separation} & \inline{int} \\
   - & \inline{angular\_separation} & String listing the unique angular separations of the merged entries & \inline{str}\\
 \inline{catalog}&\inline{catalog}&  String listing the catalogs in which the target appears& \inline{str}\\
 - & \inline{duplicate\_catalog\_flag} & Flag for duplicates during merging  & \inline{int}\\
 - & \inline{duplicate\_names} & String listing the planet name in all available input sources (only if \inline{duplicate\_catalog\_flag} is not null)&\inline{str}\\

  \inline{MismatchFlagHost}& \inline{binary\_coordinate\_mismatch\_flag} & $\star$ Flag for possible coordinate mismatches in binaries &\inline{int}\\
- & \inline{binary\_complex\_system\_flag} & $\star$ Flag for complex binary systems &\inline{int}\\
 - & \inline{coordinate\_mismatch\_flag} & Flag for possible coordinate mismatches & \inline{int}\\
 - & \inline{coordinate\_mismatch} & String labeling the coordinates for which a coordinate mismatch occurred, if any)& \inline{str}\\

 - & \inline{period\_mismatch\_flag} & Flag for potential duplicates in final catalog because of mismatching \inline{p} or \inline{a} & \inline{int}\\
 - & \inline{fallback\_merge\_flag} &
Flag for potential duplicates in final catalog because no \inline{p} or \inline{a} are available (fallback merge case) & \inline{int}\\
- & \inline{misnamed\_duplicates\_flag} & 
Flag for potential duplicates in final catalog because of a mismatch in \inline{exo-mercat\_name} when \inline{p} or \inline{a} are the same & \inline{int}\\
 - & \inline{row\_update} & Date of the last update of the row. & \inline{str}\\

  \hline  

\end{longtable}

\twocolumn

\bibliographystyle{elsarticle-harv} 
\bibliography{main}




%%\bibitem[ ()]{}

%%\end{thebibliography}
\end{document}